\documentclass[11pt,twoside]{article}

\usepackage{asp2006_jets}
\usepackage{graphicx}

\begin{document}
\title{Magnetic Field Structure in the Parsec Scale Jet of 3C\,273 from Multifrequency VLBA Observations}
\author{T. Savolainen,\altaffilmark{1,2} K. Wiik,\altaffilmark{2} E. Valtaoja,\altaffilmark{2} and M. Tornikoski\altaffilmark{3}}   
\affil{$^1$ Max-Planck-Institut f\"ur Radioastronomie, Auf dem H\"ugel 69, D-53121 Bonn, Germany}
\affil{$^2$ Tuorla Observatory, University of Turku, V\"ais\"al\"antie 20, FI-21500 Piikki\"o, Finland}    
\affil{$^3$ Mets\"ahovi Radio Observatory, Helsinki University of Technology, Mets\"ahovintie 114, FI-02540 Kylm\"al\"a, Finland}

\begin{abstract} 
We present first results from a multifrequency VLBA observations of
3C\,273 in 2003. The source was observed simultaneously at 5.0, 8.4,
15.3, 22.2, 43.2 and 86.2 GHz, and from this multifrequency data set,
spectra of 16 emission features in the parsec scale jet were carefully
constructed by using a new model-fitting based method. The measured
spectra and sizes of the emission features were used to calculate the
magnetic field density and the energy density of the relativistic
electrons in the different parts of the parsec scale jet, independent
of any equipartition assumption. We measure magnetic field density of
an order of 1 Gauss in the core. The magnetic energy density in the
core dominates over that of the relativistic electrons, while in the
downstream region our data are roughly consistent with an
equipartition. A strong gradient in the magnetic field density across
the jet width, coincident with a transverse velocity structure at
about 1.5 mas from the core, was found: the slower superluminal
component B2 on the northern side of the jet has a magnetic field
density two orders of magnitude lower than the faster southern
components B3 and B4.
\end{abstract}

\section{Introduction}   

For over 30 years, Very Long Baseline Interferometry (VLBI) has been
the prime observational method for studying compact extragalactic jets
and it has been used extensively to investigate the morphology,
kinematics, and polarization of the jets in parsec scales \citep[see
e.g.][]{zen97}. However, studies of the jets' continuum spectrum in
the VLBI scale have been hitherto surprisingly rare, even though the
synchrotron spectrum provides one of the very few available probes of
the magnetic field and particle energy density in the jet
\citep{mar87}. Pioneering observations of the jet spectrum in parsec
scales were done in the 1980s \citep{cot80,bar84,mar85,mar88}, but
before the advent of the Very Long Baseline Array in 1993, it was not
usually possible to observe the source simultaneously at several
frequencies, and the typical flux density variability of the compact
jets made it difficult to combine observations from different
epochs. VLBA's frequency agility allows nowadays practically
simultaneous multifrequency observations with a wide frequency
coverage, and this has made the studies of the continuum spectrum of
the parsec scale jets feasible
\citep[see][]{lob98,wal00,mar01,ver03,sav06a}. However, due to tricky
image alignment and uneven $(u,v)$ plane coverage between different
frequencies, extraction of the continuum spectra from a multifrequency
VLBI data set is still a non-trivial task and requires a very careful
planning, execution and analysis of the experiment. Only a very small
number of sources have been studied so far.

In the present paper, we report the first results from the spectral
analysis of multifrequency VLBA observations of the quasar 3C\,273
($z=0.158$). In Section 2, we briefly describe our new method for
extracting spectra of individual emission features in the jet from a
multifrequency VLBI data set. In Section 3, the spectra observed on
February 28, 2003 are presented, and the measured synchrotron turnover
frequencies and flux densities are used to estimate the physical
conditions in the jet, i.e. the magnetic field and the relativistic
electron energy density. The results are discussed in Section 4.

Throughout the paper, we use a cosmology with $H_0$ = 71 km s$^{-1}$
Mpc$^{-1}$, $\Omega_M$ = 0.27, and $\Omega _\Lambda$ = 0.73. This
corresponds to a linear scale of 2.7 pc mas$^{-1}$ for 3C\,273. For
the spectral index $\alpha$, we use the positive convention: $S_{\nu}
\propto \nu^{+\alpha}$.

\section{Extraction of the spectra from multifrequency VLBI data}

In the year 2003, 3C\,273 was the target of a multiwavelength campaign
organized to support the observations done with the INTEGRAL
$\gamma$-ray satellite \citep{cou03}. To complement this campaign with
imaging data, we carried out a polarimetric multifrequency monitoring
of 3C\,273 using the VLBA. These observations are described in
\citet[hereafter Paper I]{sav06b}, where a kinematical
analysis of the component motions was presented. 3C\,273 was observed
with the VLBA five times in 2003 for nine hours at each
epoch. Observations were made at six frequencies (5.0, 8.4, 15.3,
22.2, 43.2 and 86.2 GHz) and individual scans at different frequencies
were interleaved in order to obtain practically simultaneous
multifrequency data set. The calibration, imaging and self-calibration
of the data were done by the standard procedures.

In the data reduction, special attention was paid to the amplitude
calibration. The accuracy of the final flux density scale was checked
by comparing the extrapolated zero baseline flux density of the
compact calibrator source 3C\,279 at 5, 8, 22 and 43 GHz, to the
interpolated flux densities from the VLA polarization monitoring
program \citep{tay00}, and at 86\,GHz to the quasi-simultaneous data
from the SEST telescope. The accuracy turned out to be as good as
$\sim 5\%$ at the frequencies from 5 to 43\,GHz and $\sim 20 \%$ at
86\,GHz.

\begin{figure}[t]
\centering
\includegraphics[width=0.8\textwidth]{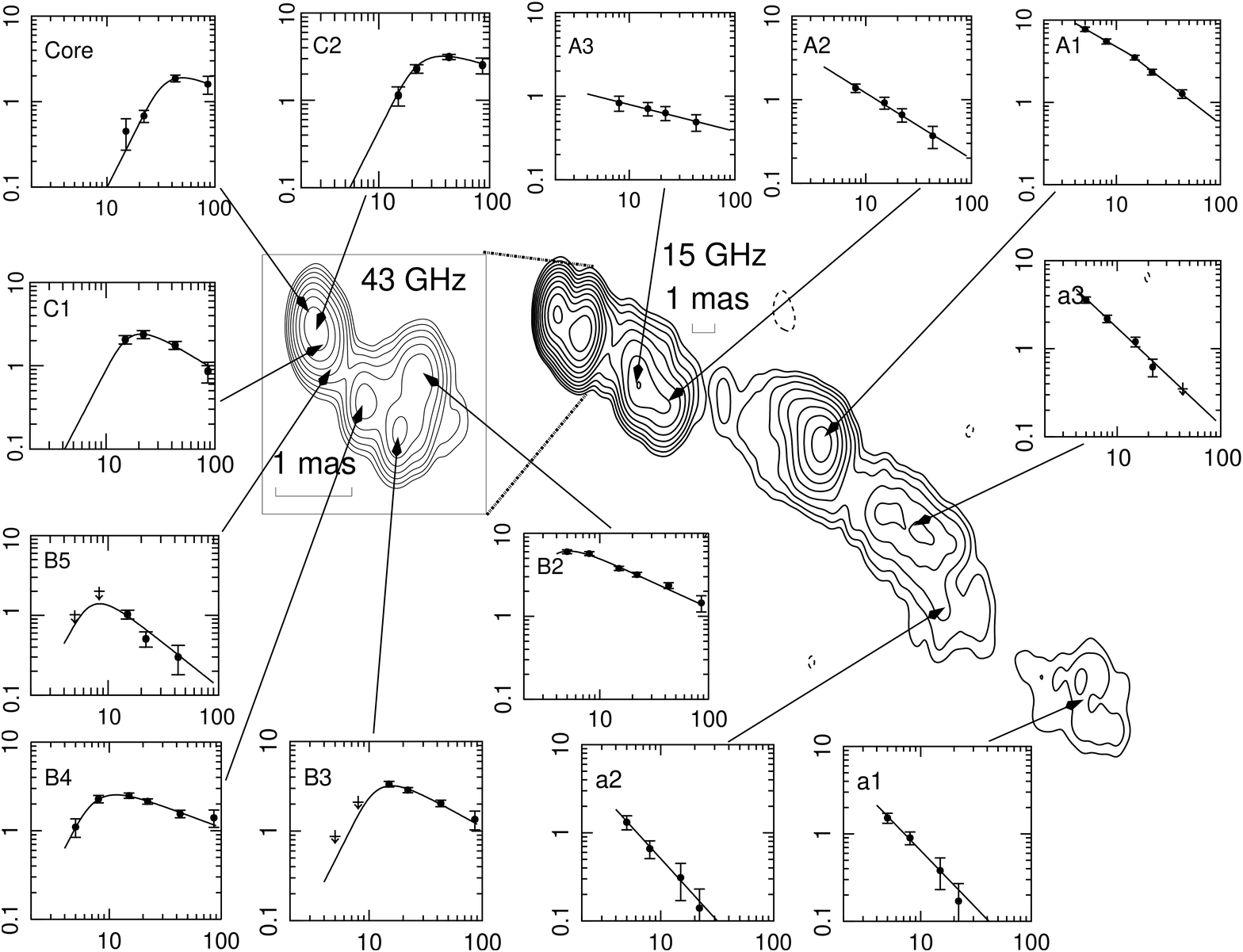}
\caption{The radio spectra of the emission features in the parsec
scale jet of 3C\,273 observed on Feb 28, 2003. The figure also shows
jet images at 15 and 43\,GHz. In the spectral plots, the x-axis is
frequency in GHz and the y-axis is flux density in Jy. The solid lines
show either a power-law or a self-absorbed synchrotron spectrum fitted
to the data. Downward arrows represent upper limits. Spectra of three
weak components (B1, a4, a5) are omitted from the figure due to lack
of space.}
\end{figure}

Because the VLBI networks are not reconfigurable like, for example,
the VLA, their $(u,v)$ plane coverage differs significantly at
different observing frequencies. Typically, in the spectral analysis
of the multifrequency VLBI data, the $(u,v)$ coverages at different
frequencies have been matched either by simply throwing away the data
at the large $(u,v)$ radii at high frequencies or by tapering the data
to a common (low) resolution. This has the consequence that a
significant amount of data are not used and much of the attainable
angular resolution is lost. A broad frequency coverage, which is
needed to reliably measure the turnover of the synchrotron spectrum,
exacerbates the problem: for a frequency coverage of $5-86$\,GHz at
the VLBA, the common range of $(u,v)$ distances between the
frequencies is only 4\% of the whole range of observed $(u,v)$ radii.

By using a model-fitting based spectral extraction method, the
above-described problem with an uneven $(u,v)$ coverage can be
significantly relieved. The idea is to use {\it a priori} knowledge of
the source structure, measured at high frequencies, to allow at lower
frequencies the derivation of sizes and flux densities of even those
emission features that have mutual separations less than the Rayleigh
limit at the given frequency. This is possible because, if we have a
template of the brightness distribution, the minimum resolvable size
is a function of the signal-to-noise ratio of the visibility data, and
for high SNR data, it can be much smaller than the Rayleigh limit
\citep[see e.g.][]{kov05}. In practice, this was done in the following
way:

1) A simple source model, consisting of two-dimensional Gaussian
   components, was formed at 43\,GHz. (In our case, since there were
   only 6 working antennas at 86\,GHz, the best angular resolution was
   obtained at 43\,GHz.)

2) This template model was transferred to 22 and 86\,GHz. The relative
   positions of the components were fixed to the values obtained from
   model-fitting at 43\,GHz, and the models were aligned by assuming
   that optically thin components have frequency independent
   positions. 

3) Model-fitting was run at each frequency letting only component
   sizes and flux densities in the transferred model to vary. If any
   two components had a separation smaller than $\sim1/5$ of the beam
   size at the given frequency, these were replaced by single
   component in order not to try to extrapolate beyond $(u,v)$ radius
   corresponding to $\sim1/5$ of the beam size, which is the typical
   uncertainty of a component position (see Paper I). If the residual
   map contained significant emission that was not accounted for by
   the transfered model, a new model component(s) was added and
   model-fitting was run again. This produced a template model for the
   next lower frequency and steps 2 and 3 were repeated for all the
   frequencies.

4) We assume that the angular size of a component varies smoothly over
   the frequencies. The sizes of the components were inspected as a
   function of frequency, and fitted with a power-law after removing
   clear outliers. The results indicated that sizes are nearly
   constant or decreasing slowly as a function of frequency (the
   power-law indeces range from -0.3 to +0.1 with typical
   uncertainties of $0.1-0.2$).

5) The angular sizes of the components were fixed to the values
   derived from the power-law fit (typically a constant), and the
   model-fit was run again at all frequencies with the component flux
   densities as the only free parameters. This resulted in the final
   spectra for the components.

\begin{table}[t]
\caption{Fitted spectral and size parameters for inner jet components}
\label{spectra_inner}
\begin{center}
\begin{tabular}{ccccccc} \tableline 
Comp. & $z_0$ & $S_{\mathrm{m}}$ & $\nu_{\mathrm{m}}$ & $\alpha$ & $a({\nu_\mathrm{m}})$ &  $\delta$ \\
 & (mas) &(Jy) & (GHz) & & (mas) & \\ \tableline
Core& 0.00 & $1.9\pm0.2$ & $50\pm10$ & $-0.6\pm0.4$ & $0.07\pm0.01$  & -- \\
C2  & 0.15 & $3.2\pm0.1$ & $38\pm1$  & $-0.4\pm0.1$ & $0.09\pm0.02$  & $5.5\pm1.9$ \\
C1  & 0.30 & $2.4\pm0.2$ & $21\pm1$  & $-0.8\pm0.2$ & $0.10\pm0.03$  & -- \\
B5  & 0.54 & $1.4\pm0.2$ & $8\pm1$   & $-1.1\pm0.3$ & $\approx0.16$  & -- \\
B4  & 1.07 & $2.5\pm0.1$ & $11.5\pm0.6$ & $-0.5\pm0.1$ & $0.22\pm0.03$ & $8.6\pm3.2$ \\
B2  & 1.31 & $6.1\pm0.3$ & $5.2\pm0.7$ & $-0.6\pm0.1$ & $0.28\pm0.06$  & $7.1\pm2.5$\\
B3  & 1.54 & $3.2\pm0.1$ & $15.8\pm0.7$ & $-0.7\pm0.1$ & $0.23\pm0.03$ & $4.2\pm1.7$ \\
B1  & 1.66 & $0.6\pm0.2$ & $11\pm5$ & $-0.2\pm0.2$ & $0.10\pm0.03$  & -- \\ \tableline
\end{tabular}
\end{center}
\end{table}

The assumption of a smoothly varying component angular size allows us
to use only a small number of parameters in the final model, resulting
in robust estimates of the component spectra even in the size scales
significantly smaller than the resolving beam. A more detailed
description of our spectral extraction method together with an error
analysis is given in Savolainen et al. (A\&A, submitted).

\section{Component spectra and derived magnetic field densities}

\begin{figure}[t]
\centering
\includegraphics[angle=-90,width=0.58\textwidth]{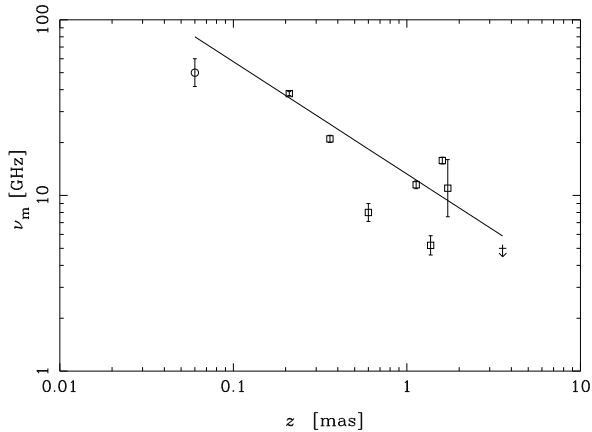}
\caption{The synchrotron peak frequency as a function of distance
along the jet. Since the total radio spectrum of 3C\,273 steepens
above $\sim 60$\,GHz, the jet cannot stay self-similar beyond the
point where $\tau=1$ surface is at this frequency. The mm-core we
observe has $\nu_\mathrm{m}=50\pm10$\,GHz and we do not register any
core shift between 43 and 86 GHz images within the accuracy of
0.02\,mas. Hence, we conclude that our core component is within
0.06\,mas (half the $FWHM$ of the core times a geometrical correction
factor of 1.8) from the point where the self-similarity breaks up. We
measure the distance $z$ with respect to this point
($z=z_0+0.06$\,mas). The open circle corresponds to the core, while
the downward arrow shows the upper limit of $\nu_\mathrm{m}$ at $z >
3.5$\,mas. The solid line is a power-law fit to the data:
$\nu_\mathrm{m} \propto z^{-0.6\pm0.1}$.}
\end{figure}

The final spectra are presented in Fig.~1. The components within
2\,mas from the core show a spectral turnover at our frequency
range. We have fitted the spectra of these components with a function
describing self-absorbed synchrotron radiation emitted by electrons
having a power-law energy distribution in a homogeneous magnetic
field. From the fits, values of the turnover frequency,
$\nu_\mathrm{m}$, the maximum flux density reached at the turnover
frequency, $S_\mathrm{m}$, and the optically thin spectral index,
$\alpha$, were derived. These values are given in Table~1 together
with the components' distances from the core, $z_0$, their sizes, $a$,
at $\nu_\mathrm{m}$, and their Doppler factors, $\delta$, which were
measured in Paper~I.\footnote{There was an error in the Doppler factor
reported for component B2 in Paper~I. Table~1 contains the correct
value. This value of Doppler factor, together with
$\beta_{\mathrm{app}}=6.5\pm0.6$, yields a Lorentz factor of
$6.6\pm1.4$ and a viewing angle of $8.1\pm3.0\deg$ for B2.}

In Fig.~2 we have plotted $\nu_\mathrm{m}$ as a function of distance
along the jet, $z$, in a logarithmic scale. The peak frequency seems
to decrease rather steadily as a function of $z$. A simple power-law
fit to the peak frequencies gives $\nu_\mathrm{m} \propto
z^{-0.6\pm0.1}$, at least within 2\,mas from the core. This confirms
the compound nature of 3C\,273's flat radio spectrum: components
become self-absorbed at progressively lower frequencies as they move
out along the jet creating an apparently flat radio spectrum up to
$60-90$\,GHz \citep[see Fig.~8 in][]{cou98}.

\begin{figure}[t]
\centering
\includegraphics[width=0.58\textwidth]{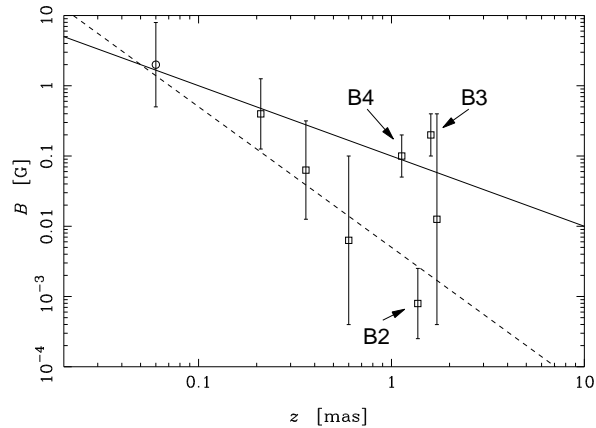}
\caption{Magnetic field density of the jet components as a function of
 distance along the jet. The open circle corresponds to the core
 component. The solid and dashed lines show examples of a power-law
 dependence $B\propto z^b$ with $b=-1$ and $b=-2$,
 respectively.}
\end{figure}

Adopting the standard synchrotron theory and assuming that emission
features are uniform and spherical, we have used the observed
component spectra -- together with the measured sizes and Doppler
factors -- to estimate the magnetic field density, $B$, the electron
energy distribution normalization factor, $N_0$, and the relativistic
electron energy density, $U_\mathrm{re}$, of the components in
Table~1. The formulae are given in \citet{mar87}. Since we have {\it
measured} $\nu_\mathrm{m}$, $S_\mathrm{m}$, $\delta$ and $a$, we can
calculate $B$ and $N_0$ separately without having to assume
equipartition conditions. The component size inserted in the formulae
is 1.8 times the $FWHM$ of the Gaussian component \citep{mar87}. Since
$\delta$ was determined only for components C2, B2, B3 and B4 in
Paper~I, we have used an average value $\langle \delta \rangle=5.6$
for the rest of the components. The very strong non-linear dependence
of $B$, $N_0$, and $U_\mathrm{re}$ on the observed quantities makes a
linear approximation of the error propagation invalid, and we have
employed a Monte Carlo approach to calculate the uncertainties. The
results are listed in Table~2.

\section{Discussion}

Fig.~3 shows $B$ as a function of $z$, in a logarithmic scale. The
core has magnetic field density of an order of 1\,Gauss, which is
compatible with the values derived from infrared and optical
variability (Courvoisier et al. 1988). Generally, the magnetic field
becomes weaker as we move out along the jet. However, there is a large
discrepancy in $B$ between components B2 and B3, both having about the
same distance from the core. Component B2, which is located
$\approx0.6$\,mas north of B3, has two orders of magnitude smaller $B$
than component B3 (or B4). The kinematic analysis presented in Paper~I
already showed that there is a significant velocity gradient across
the jet width at the same location: the northern component B2 has a
bulk Lorentz factor $\Gamma=6.6\pm1.4$, while the southern components
B3 and B4 have $\Gamma=17\pm7$ and $\Gamma=18\pm8$, respectively. The
magnetic field and velocity gradients across the jet width are thus
coincident: the slower northern component B2 has significantly smaller
$B$ than the faster southern components B3 and B4. Are we seeing a
spine/sheath structure in 3C\,273 (with B3 and B4 corresponding to a
fast spine and B2 to a slower layer), where the magnetic field density
decreases from the high jet-axis value to the lower value in the
sheath? An argument against this interpretation is the absence of
emission from the jet layer south of the spine. However, if the
magnetic field in the layer has a large enough helical component,
emission asymmetries can arise \citep{roc07}.

\begin{table}[t]
\caption{Physical parameters of the emission regions}
\label{phys_par_table}
\begin{center}
\begin{tabular}{ccccccc} \tableline
Comp. & $z_0$ & $\log_{10}(B)$ & $\log_{10}(N_0)$ & $\log_{10}(U_{\mathrm{re}})$
& $\log_{10}(U_{\mathrm{B}} / U_{\mathrm{re}})$ \\
  & (mas) & (Gauss) & (erg$^{-2\alpha}$ cm$^{-3}$) & (erg cm$^{-3}$) & \\ \tableline
Core & 0.00 & $+0.3\pm0.6$ & $-6.9\pm1.8$  & $-5.0\pm1.2$ & $+4.2\pm1.6$ \\
C2   & 0.15 & $-0.4\pm0.5$ & $-4.2\pm1.2$  & $-3.8\pm1.3$ & $+1.7\pm1.6$ \\
C1   & 0.30 & $-1.2\pm0.7$ & $-6.4\pm1.9$  & $-3.7\pm1.6$ & $-0.1\pm2.1$ \\
B5   & 0.54 & $-2.2\pm1.2$ & $-7.6\pm3.6$  & $-3.7\pm2.7$ & $-2.1\pm3.6$ \\
B4   & 1.07 & $-1.0\pm0.3$ & $-6.1\pm1.1$  & $-5.2\pm1.1$ & $+1.9\pm1.3$ \\
B2   & 1.31 & $-3.1\pm0.5$ & $-3.5\pm1.4$  & $-2.1\pm1.3$ & $-5.5\pm1.6$ \\
B3   & 1.54 & $-0.7\pm0.3$ & $-7.2\pm1.2$  & $-5.0\pm1.0$ & $+2.3\pm1.2$ \\
B1   & 1.66 & $-1.9\pm1.5$ & $-2.2\pm2.3$  & $-2.2\pm2.6$ & $-3.0\pm4.0$ \\ \tableline
\end{tabular}
\end{center}
\end{table}

An average value of $U_\mathrm{re}$ within 2\,mas from the core is
about $10^{-4}$\,erg\,cm$^{-3}$. The core has significantly more
energy in the magnetic field than in the radiating particle
population, while component B2, on the other hand, is heavily particle
dominated. The other components are more or less compatible with a
rough equipartition between $U_\mathrm{B}$ and $U_\mathrm{re}$. In
addition to magnetic field and electrons, there will be energy stored
in cold protons, if the jet is composed of electron-proton plasma. If
one cold proton per a relativistic electron is assumed and the jet is
required to be rest mass dominated beyond the core instead of being
Poynting flux dominated ($U_\mathrm{p} > U_\mathrm{B}$), components
C2, B3 and B4 must have the low energy cut-off of the relativistic
electron distribution $\la 10$.

\acknowledgements 

This work was partly supported by the Finnish Cultural Foundation
(TS), and by the Academy of Finland grants 74886, 210338, and
120516. The VLBA is a facility of the National Radio Astronomy
Observatory, operated by Associated Universities, Inc., under
cooperative agreement with the U.S. National Science Foundation.

\end{document}